       \newcommand{\nl}{\nonumber \\}
       \newcommand{\beq}{\begin{equation}}
       \newcommand{\eeq}{\end{equation}}
       \newcommand{\beqa}{\begin{eqnarray}}
       \newcommand{\eeqa}{\end{eqnarray}}
       \newcommand{\beqas}{\begin{eqnarray*}}
       \newcommand{\eeqas}{\end{eqnarray*}}
       
       \newcommand{\bB}{{\mathbf B}}

       \newcommand{\bE}{{\mathbf E}}

\documentclass[12pt]{article}
\usepackage[dvips]{graphicx}\usepackage[dvips]{graphics}
\usepackage{enumerate}
\usepackage{psfrag}
\usepackage{color}
\begin{document}
\begin{center}
{\large \bf Analytic, quasineutral, two-dimensional Maxwell-Vlasov
equilibria } \vspace{3mm}

{\large  G. N. Throumoulopoulos$^1$, H. Tasso$^2$} \vspace{2mm}

{\it
$^1$University of Ioannina, Association Euratom - Hellenic Republic,\\
 Section of Theoretical Physics, GR 451 10 Ioannina, Greece}
\vspace{2mm}

 { \it  $^2$Max-Planck-Institut f\"{u}r Plasmaphysik, Euratom
Association,\\
 D-85748 Garching, Germany }
\end{center}
%
%
\vspace{2mm}
\begin{center}
{\bf Abstract}
\end{center}
 \noindent

 Two-dimensional Maxwell-Vlasov equilibria with finite electric fields,
 axial (``toroidal")  plasma flow and isotropic pressure are constructed
  in plane geometry by using the quasineutrality
 condition to express the electrostatic potential in terms of the
 vector potential. Then for Harris-type distribution functions,  Ampere's
 equation  becomes of Liouville type and can be solved
 analytically. As an example, a periodic  ``cat-eyes" steady state
 consisting of a row of magnetic islands
  is presented. The method can be extended to
 (toroidal) axisymmetric equilibria.

\newpage
%

Equilibrium is      the starting point
 for stability and transport studies of astrophysical and laboratory plasmas.
 In the framework of collisionless kinetic  theory, equilibrium states
 should be constructed as self consistent solutions of Vlasov and
 Maxwell equations. To this end,  the
 knowledge of constants of motion for the particles in the
 continuum approximation (microfluids) is of crucial importance because then the general
 solution of Vlasov equation can be written as an arbitrary function of the
 complete set of constants of motion. This is feasible only for
 one-dimensional equilibria,  i.e. in this case the three constants of motion
 are the energy ($H_s=(1/2)m_s v^2 +q_s \Phi(y)$) and the two canonical
 momenta ($p_{xs}=m_s v_x+q_s A_x(y)$ and $p_{zs}=m_s v_z+q_s A_z(y)$);
  consequently, the distribution functions are of the form
 $f_s=f_s(H_s, p_{xs}, p_{zs})$. Here, ($x,y,z$) are Cartesian
 coordinates,
 $v^2=v_x^2+v_y^2+v_z^2$, $\Phi(y)$ the electrostatic
 potential,
 $A_x(y)$ and $A_z(y)$  the components of the
 vector potential, and the subscript $s$ denotes the particle
 species.
 Unlikely, for two-dimensional equilibria the complete
 set of constants of motion is missing, i.e. only the energy $H_s$
 and the momentum $p_{zs}=m_s v_z+q_s A_z(x,y)$ conjugate to the ignorable
 coordinate $z$ are known out of
 the four constants of motion. A good number of solutions for
 one-dimensional
 \cite{Cha}-\cite{HaNe1}
 and two-dimensional \cite{CeMo}-\cite{Su}  equilibria were constructed
  on the basis of modified Maxwellian distribution functions of the forms $f_s=\exp(-\beta_s H_s)g_s(p_{xs},
  p_{zs})$ and $f_s=\exp(-\beta_s H_s)g_s(p_{zs})$, respectively, with $g_s$ arbitrary functions
   of the conserved momenta and
  $\beta_s=1/(k_B T_s)$.
  These equilibria concern neutral plasmas in connection  with a
  special set of distribution functions such that it holds  (for one-dimensional equilibria)
   $N_i(A_x, A_z)=N_e(A_x, A_z)$, where $N_i$ ($N_e$) is the
  ion (electron) density \cite{Cha}; viz. in addition to the usual quasineutrality
  condition  it was assumed that $N_i(A_x,A_z)$ is the
  same function of $A_x$ and $A_z$ as $N_e(A_x,A_z)$, thus leading
  to a vanishing electrostatic potential. On physical grounds this
  additional assumption is oversimplifying because it ignores
  the mass difference of ions and electrons. Also, finite electric
  fields associated with macroscopic plasma (ion) flows
  are important in laboratory fusion
  plasmas for the transitions from low to high confinement modes of operation.

  Aim of the present study is to construct analytically
  a class of quasineutral, two dimensional Maxwell-Vlasov
  equilibria. This is accomplished by   employing the quasineutrality
  condition (without the additional assumption of functionally
  identical ion and electron densities) to express $\Phi$ as a
  function of $A_z(x,y)$. A similar method was employed by Mynick
  and coauthors \cite{MySh} to construct numerically by an iteration
  algorithm
   one-dimensional quasineutral
  equilibria. Also, the method was  reviewed recently in  Sec. II of Ref. \cite{HaNe2}.

  We consider a plasma   of
  electrons  and protons at equilibrium with a current density in the axial (``toroidal") $z$-direction.
  Consequently, the
  vector potential has a single component $A_z(x,y)$. In addition
  to the ``poloidal" magnetic field with components $B_x(x,y)$ and
  $B_y(x,y)$ associated  with $A_z$,
  we include  for stabilizing reasons a constant axial
   magnetic field $B_{z0}$ which otherwise does not affect the
  equilibrium. Furthermore, we employ Harris-type distribution
  functions,
  \beq
  f_s(H_s, p_{zs})= n_{0s}\left(\frac{m_s \beta_s}{2\pi}\right)^{3/2}\exp(-\beta_s H_s)\exp(\beta_s V_{zs} p_{zs}),
                                                      \label{1}
  \eeq
  where $V_{zs}$ are constant average  (fluid)
  velocities and $n_{0s}$ reference densities corresponding to Maxwellian distribution functions ($V_{zs}=0$).
  Using the quasineutrality condition, $N_i=N_e$, where
  $$ N_s=n_{0s}\left(\frac{m_s \beta_s}{2\pi}\right)^{3/2}
         \exp(-q_s\beta_s\Phi)\int_{-\infty}^{\infty} \exp\left(-\beta_s m_s v^2/2\right)
         \exp(\beta_s V_{zs} p_{zs}) \, d^3 v$$
  (with $q_i=e$  and $q_e=-e$), the electrostatic potential can be
  expressed in terms of $A_z(x,y)$ as
  \beq
  \Phi(A_z)=\log\left\lbrack \frac{n_{0i}}{n_{0e}}\exp
           \left( A_z e V_{ze} \beta_e-\frac{1}{2}m_e
           V_{ze}^2\beta_e +  A_z e  V_{zi} \beta_i+\frac{1}{2}m_i
           V_{zi}^2\beta_i\right)\right\rbrack^{\frac{1}{e(\beta_e+\beta_i)}}.
                                            \label{2}
  \eeq
  Note that for (1) all the integrations of interest in the velocity space can
  be performed analytically.
 Using (\ref{1}) and (\ref{2}) one finds for the current density
  ($j_z=\sum_{s}q_s\int_{-\infty}^{\infty} v_z f_s d^3 v$):
  \beqa
  \lefteqn{j_z(A_z)=
  e
  n_{0e}^{\frac{\beta_i}{\beta_e+\beta_i}}n_{0i}^{1-\frac{\beta_i}{\beta_e+\beta_i}}\left(V_{zi}-V_{ze}\right)}
  \nl
 & & \exp\left\lbrack \frac{1}{2}V_{zi}\left(2 A_z e + m_i
      V_{zi}\right)\beta_i-\frac{\beta_i\left(A_z e
      V_{ze}\beta_e-\frac{1}{2}m_e V_{ze}^2 \beta_e+A_z e
      V_{zi}\beta_i+\frac{1}{2}m_i V_{zi}^2
      \beta_i\right)}{\beta_e+\beta_i}\right\rbrack. \nl
                                                      \label{3}
  \eeqa
Therefore, Ampere's equation ($\nabla^2 A_z(x,y)=-\mu_0 j_z(A_z)$)
 assumes a Liouville-type form
 \beq
 \nabla^2 A_z= a\exp(b A_z)
                                                \label{4}
 \eeq
 where
 \beqa
 a&=&\mu_0 \left( V_{ze}-V_{zi}\right) e
 n_{0e}^{\frac{\beta_i}{\beta_e+\beta_i}}
    n_{0i}^{\frac{\beta_e}{\beta_e+\beta_i}}
    \exp\left\lbrack \frac{\left(m_e V_{ze}^2 + m_i
    V_{zi}^2\right) \beta_i \beta_e}{2\left(\beta_e
    +\beta_i\right)}\right\rbrack, \label{4a} \\
 b&=&\frac{e \left(
 V_{zi}-V_{ze}\right)\beta_e\beta_i}{\beta_e+\beta_i}.
                                               \label{4b}
 \eeqa
 The equilibrium for $V_{zi}=0$ is static, viz. only the electrons are
 in non-thermal motion to produce the axial current. For $V_{zi}\neq 0$,
 however,
 there is a constant ion-fluid axial velocity,
 $\int_{-\infty}^{\infty} v_z f_i d^3 v/N_i= V_{zi}$, and the ion
 motion contributes to $j_z$. It may be noted here that for distribution functions of the form
 $f_s(H_s, p_{zs})$ it is not possible to create poloidal
 plasma
 velocities because of the   two missing constants of
 motion. Even the third constant of motion found in  Ref. \cite{TaTh} near the magnetic axis
 does not help to this end    because poloidal flows  vanish on axis.  For
 the equilibrium constructed here the pressure is isotropic, i.e. the pressure
 tensor,
 $$P_{ij}= \sum_{s} m_z \int_{-\infty}^{\infty} (v_i -\langle v_i \rangle_s)(v_j -\langle v_j \rangle_s) f_s d^3 v,
 \ \ i,j=x,y,z,
 $$
 is diagonal
 with $P_{xx}=P_{yy}= P_{zz}\equiv P$
 (see also Fig. 3).
  For $V_{zi}=V_{ze}$ it follows that the current density vanishes
 ($a=b=0$) and (\ref{4}) reduces to Laplace equation. Therefore, $A_z$ can not
 be constant on any closed curve in the ($x,y$) plane without
 being constant in the region within this curve.
 Consequently, the electrostatic potential
 $\Phi$ is constant too in this region because of (\ref{2})
  and the distribution functions become spatially
  uniform;  hence,  one recovers the well-known equilibrium solution
  of the Maxwell-Vlasov equations for which all quantities are
  homogeneous. No ``confined solution" is possible in this case.
  Also, it is noted here that for $V_{zi}=V_{ze}$ solution (\ref{5a}) below,
  though pertinent to an unbounded plasma, becomes singular ($\tilde{A}_z\rightarrow
  \infty$).

  Introducing  dimensionless  quantities ($\xi=x/L$, $\eta=y/L$,
  $\tilde{A}_z=A_z/(B_{z0}L)$, $\tilde{a}=a L/B_{z0}$, $\tilde{b}=b B_{z0} L$ with $L$ a
 length scaling parameter),
  the general solution of (\ref{4}) is given by \cite{Li,Cle}
  \beqa
  \tilde{A}_z(x,y)&=&\frac{\chi\left|\tilde{a} \tilde{b}\right|-
  \tilde{a}\tilde{b}\log\left|\tilde{a} \tilde{b}\right|}{\tilde{a} \tilde{b}^2}, \nl
  \chi&=&\log\left\lbrack\frac{\left(u^2+v^2+1\right)^2}{8
  \left(u_\xi^2+u_\eta^2\right)}\right\rbrack,
                                            \label{5}
  \eeqa
  where $u(\xi,\eta)$ and $v(\xi,\eta)$ are real conjugate functions
  resulting from $w(\xi+i \eta)=u+i v$, with $w$ a differentiable
  arbitrary  generating function. As an example of complete
  equilibrium construction  we consider here the function
  $$
  w(\zeta)=\sqrt{\frac{1+\epsilon}{1-\epsilon}}
  \tan\left(\frac{\zeta}{2}\right)
  $$ with
  $\zeta= \xi + i \eta$, and $\epsilon $ a parameter such that $|\epsilon|\leq
  1$.
  For this choice of $w$, (\ref{5})
  acquires  the form
  \beq
  \tilde{A}_z(\xi,\eta)=\log\left\{\frac{2 (1-\epsilon^2)}{|\tilde{a}
  \tilde{b}|\left\lbrack\cosh\left(\xi\right)-\epsilon
  \cos\left(\eta \right)\right\rbrack^2} \right\}^{1/\tilde{b}}.
                                               \label{5a}
  \eeq
  The function $\tilde{A}_z$ labels the magnetic surfaces.
 The equilibrium configuration shown in Fig. 1
 consists  of an infinite row of identical periodic islands known as ``cat-eyes"
 (see for example  Ref.   \cite{ThTa}). The islands have magnetic axes
 at $\xi_a=2 k \pi$, $\eta_a=0$ and a separatrix with $x$-points
 at $\xi_x=(2k+1) \pi $, $\eta_x=0$ where $k$ an integer. The ordinates   of the separatrix
 are located at $\xi=\xi_a$, $\eta=\eta_s=\pm  \ \mbox{arctanh}\left(1+2 \epsilon\right)$ (see Fig. 1).
    The
 equilibrium
  has the following  free parameters:  $n_{0s}$, $\beta_s$,
  $V_{zs}$ ($s=e,i$),
  $B_{z0}$, $\epsilon$, and  $L$.
  The dependent parameters $a$ and $b$ (Eqs. (\ref{4a} and (\ref{4b})) relate
  features of the distribution function
    to macroscopic equilibrium characteristics (Eq. (\ref{5a})).  For $V_{zi}\neq V_{ze}$ the physical
    quantities
  ($\bB$, $\bE=-(d\Phi/dA_z)\nabla A_z$, $j_z$ and  $P$) are everywhere regular and
  vanish as $y$ tends to infinity  except for $E_y$ which in this limit
  approaches a finite
   value.
  Profiles of $E_x$ and $E_y$ are shown in Fig. 2 for the following fusion
   relevant values of the free parameters:  $n_{0i}=n_{0e}=10^{19} m^{-3}$, $k_B T_i=k_B T_e=1 keV$,
  $V_{zi}=10^4 m/sec$, $V_{ze}=10 V_{zi}$, $B_{z0}=1 T$, $L=1 m$, and $\epsilon=0.6$.
   Also, the $y$-profiles of $j_z$ and  $P$ have an extremum on the magnetic axis (Fig. 3). For
  $\epsilon=0$ the configuration becomes one-dimensional;
  this is as an extension of the Harris sheet equilibrium \cite{Ha} (usually
  employed as initial state in reconnection studies) with finite
  $\bf E$ and constant axial velocity.

  Quasineutral equilibria with sheared axial flow which may be more pertinent to
  the improved confinement modes can be constructed by the alternative distribution functions
  \beq
  f_s(H_s, p_{zs})= n_{0s}\left(\frac{m_s \beta_s}{2\pi}\right)^{3/2}\exp(-\beta_{z} H_s)
                    \exp\left(\frac{\beta_{zs}  p_{zs}^2}{2 m_s}\right),
                                                      \label{6}
  \eeq
  with $\beta_s$ and $\beta_{zs}$ constants.
  A similar procedure then leads to $A_z$-dependent average
  axial velocities:
  $$
  \frac{\int_{-\infty}^{\infty} v_z f_s d^3 v}{N_s}=\frac{q_s
  \beta_{zs}}{m_s(\beta_s-\beta_{zs})} A_z(x,y),
  $$
  and Ampere's equation assumes the form
  \beq
  \nabla^2 A_z=a_1 A_z\exp\left(b_1 A_z^2\right),
                                               \label{7}
  \eeq
  where the parameters $a_1$ and $b_1$ are known functions of
  $n_{0s}$, $\beta_s$ and $\beta_{zs}$. Eq. (\ref{7}), higher
  nonlinear than (\ref{4}),
   should   be solved numerically.

  In summary, using the quasineutrality condition to express the
  electrostatic potential in terms of the vector potential and Harris-type distribution
  functions (Eq. (\ref{1})) we have constructed a class of plane,
  two-dimensional Maxwell-Vlasov equilibria with finite electric
  fields, constant axial plasma velocity and isotropic pressure.
  The equilibrium was  exemplified by the cat-eyes solution.
  Equilibria with sheared axial flow can be derived by alternative
  distribution functions, e.g. (\ref{6}). The method can also be applied
  in (toroidal)  axisymmetric and helically symmetric geometries.



 One of the authors (GNT) would like to thank Prof. H. Weitzner for useful discussions.
 Part of this work was conducted during a visit of  GNT
 to the Max-Planck-Institut f\"{u}r Plasmaphysik, Garching.
 The hospitality of that Institute is greatly appreciated.
 This work was performed  within the participation of the
University of Ioannina in the Association Euratom-Hellenic
Republic, which is supported in part by the European Union and by
the General Secretariat of Research and Technology of Greece. The
views and opinions expressed herein do not necessarily reflect
those of the European Commission.
 \newpage

  \newpage
\vspace*{-1cm}
 \begin{center}
 {\large \bf  Figure captions}
 \end{center}

 \noindent
 Fig. 1: \ $\tilde{A}_z$-lines of the cat-eyes solution (\ref{5})
 as intersections of the magnetic surfaces with the
 poloidal plane.
 \vspace{0.3cm}

\noindent
 Fig. 2: \ Profiles of the electric field components $E_x$ and $E_y$ associated with the
           cat-eyes solution (\ref{5}). The profiles $E_x(y)$ and
           $E_y(x)$ have chosen at $x/L=\pi/2$ and $y/L=0.5$,
           respectively, because $E_x(x=0,y)=E_y(x, y=0)\equiv 0$.
\vspace{0.3cm}

\noindent
 Fig. 3: \  $y$-profiles at $x=0$ of the current density, $j_z$, and
the pressure,  $P$, associated
            with the cat-eyes solution (\ref{5}).

 \newpage

 \begin{center}
 {\large \bf  List of Figures}
 \end{center}
 \begin{figure}[!h]
 \begin{center}
 \includegraphics[scale=0.8]{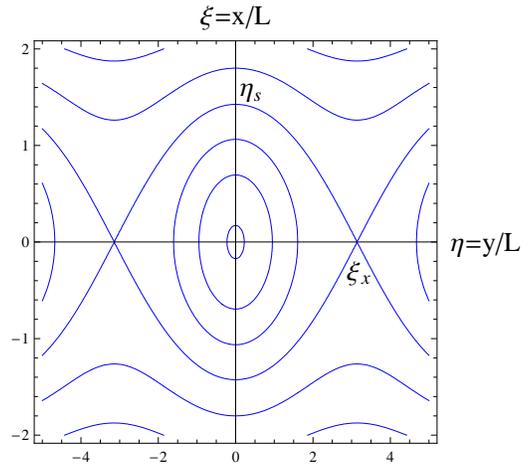}
 \caption{$\tilde{A}_z$-lines of the cat-eyes solution (\ref{5})
 as intersections of the magnetic surfaces with the
 poloidal plane.}
 \label{fig:1a}
 \end{center}
 \end{figure}
  \vspace{-0.8cm}
 \begin{figure}[!h]
 \begin{center}
 \includegraphics[scale=0.8]{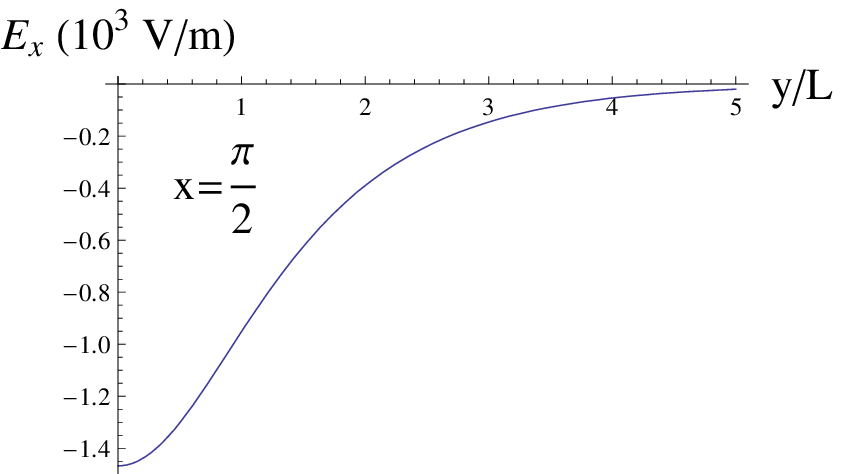}
 \includegraphics[scale=0.8]{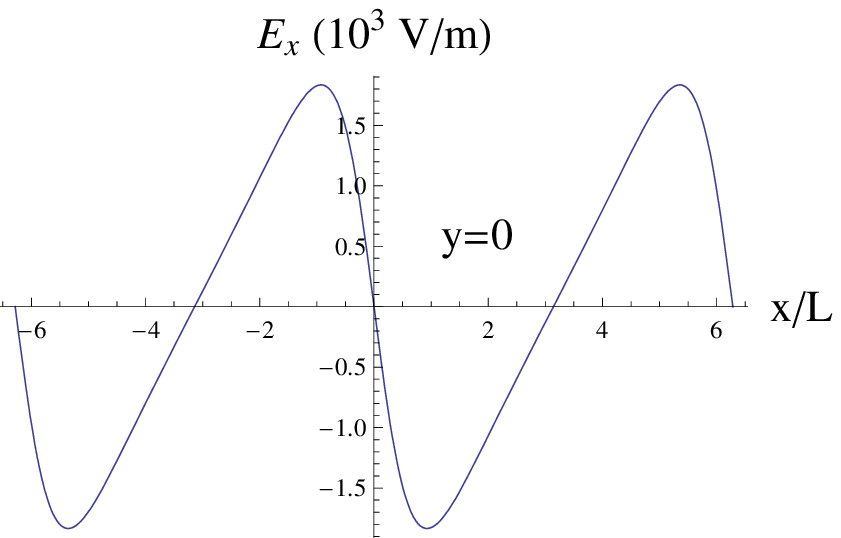}
 \includegraphics[scale=0.8]{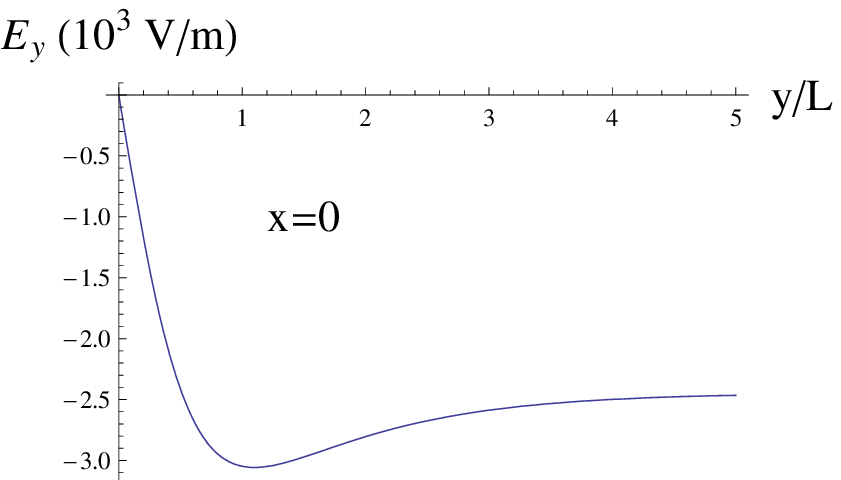}
 \includegraphics[scale=0.8]{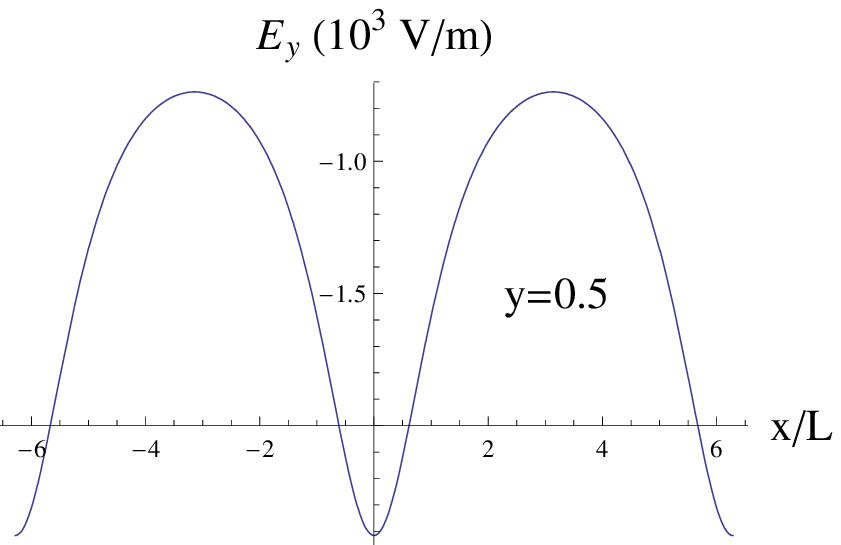}
 \caption{Profiles of the electric field components $E_x$ and $E_y$ associated with the
           cat-eyes solution (\ref{5}). The profiles $E_x(y)$ and
           $E_y(x)$ have chosen at $x/L=\pi/2$ and $y/L=0.5$,
           respectively, because $E_x(x=0,y)=E_y(x, y=0)\equiv 0$.}
 \label{fig:1b}
 \end{center}
 \end{figure}

 \begin{figure}[!h]
 \begin{center}
 \includegraphics[scale=0.8]{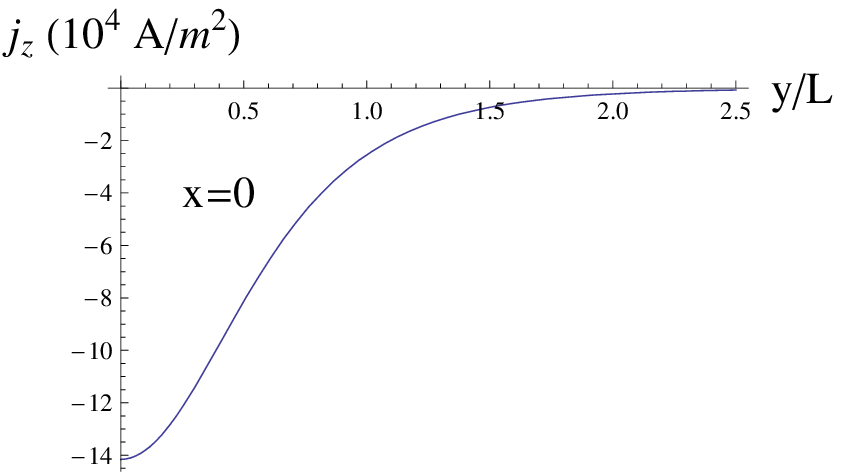}
  \includegraphics[scale=0.8]{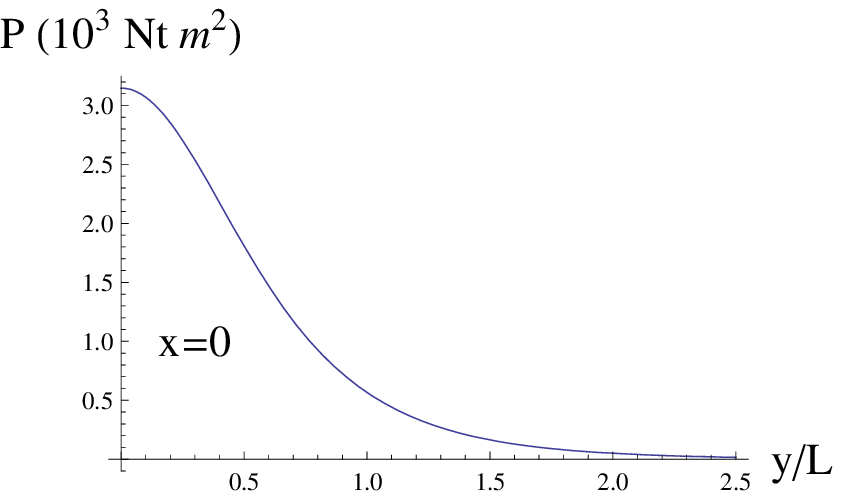}
 \caption{ $y$-profiles at $x=0$ of the current density, $j_z$, and
the pressure,  $P$, associated
            with the cat-eyes solution (\ref{5}).}
 \label{fig:2a}
 \end{center}
 \end{figure}


\begin{thebibliography}{99}
 \bibitem{Cha} Paul J. Channell, Phys. Fluids {\bf 19}, 1541 (1976).
 \bibitem{Ma} Swadesh M. Mahajan, Phys. Fluids B  {\bf 1}, 43 (1989).
 \bibitem{AtPe} N. Attico, F. Pegoraro, Phys. Plasmas {\bf 6}, 767 (1999).
 \bibitem{Mot1} F. Mottez, Phys. Plasmas {\bf 10}, 2501 (2003).
 \bibitem{Mot2} F. Mottez, Ann. Geophys. {\bf 22}, 3033 (2004).
 \bibitem{MoPe} C. Montagna and F. Pegoraro, Phys. Plasmas {\bf 14}, 042103 (2007).
 \bibitem{HaNe1} Michael G. Harrison and Thomas Neukirch   PRL {\bf 102}, 135003 (2009).
 \bibitem{CeMo} F. Ceccherini,  C. Montagna, F. Pegoraro, G. Cicogna, Phys. Plasmas 12, 052506 (2005).
 \bibitem{SuSh} Akihiro Suzuki and Toshikazu Shigeyama,  Phys. Plasmas 15, 042107 (2008).
 \bibitem{Su} Akihiro Suzuki   Phys. Plasmas 15, 072107 (2008).
 \bibitem{MySh} Harry E. Mynick, William M. Sharp, and A. N.
                Kaufman, Phys. Fluids. {\bf 22}, 1478 (1979).
 \bibitem{HaNe2} Michael G. Harrison and Thomas Neukirch   Phys. Plasmas {\bf 16}, 022106 (2009).
 \bibitem{TaTh} H. Tasso and G. N. Throumoulopoulos,   J. Phys. A: Math. Theor.  {\bf 40}, F631 (2007).
 \bibitem{Li} J. Liouville,   J. Math.  {\bf 18} (1), 71 (1853).
 \bibitem{Cle} Roberto A. Clemente,   Int. J. Math. Educ. Sci. Technol.  {\bf 23}, 620 (1992).
 \bibitem{ThTa} G. N. Throumoulopoulos, H. Tasso, G. Poulipoulis, J. Phys. A: Math. Theor.
 {\bf 42}, 335501 (2009).
 \bibitem{Ha} E. G. Harris,  Nuovo Cimento {\bf 23}, 115 (1962).

 \end{thebibliography}
 \end{document}